\documentclass{article}

\usepackage{PRIMEarxiv}

\usepackage[utf8]{inputenc} 
\usepackage[T1]{fontenc}    
\usepackage{hyperref}       
\usepackage{url}            
\usepackage{booktabs}       
\usepackage{amsfonts}       
\usepackage{nicefrac}       
\usepackage{microtype}      
\usepackage{lipsum}
\usepackage{fancyhdr}       
\usepackage{graphicx}       
\usepackage{subcaption}
\usepackage{amsmath}
\usepackage{tabularx}
\usepackage{cellspace}
\graphicspath{{media/}}     

\title{Joint Design of 5' Untranslated Region and Coding Sequence of mRNA}

\author{
  Yang Liu, Jie Gao, Xiaonan Zhang, Xiaomin Fang \\
  PaddleHelix Team\thanks{paddlehelix.baidu.com} \\
  Baidu Inc. \\
}

\begin{document}
\maketitle

\begin{abstract}
Messenger RNA (mRNA) vaccines and therapeutics are emerging as powerful tools against a variety of diseases, including infectious diseases and cancer. The design of mRNA molecules, particularly the untranslated region (UTR) and coding sequence (CDS) is crucial for optimizing translation efficiency and stability. Current design approaches generally focus solely on either the 5' UTR or the CDS, which limits their ability to comprehensively enhance translation efficiency and stability. To address this, we introduce LinearDesign2, an algorithm that enables the co-design of the 5' UTR and CDS. This integrated approach optimizes translation initiation efficiency (TIE), codon adaptation index (CAI), and minimum free energy (MFE) simultaneously. Comparative analyses reveal that sequences designed by LinearDesign2 exhibit significantly higher TIE than those designed by LinearDesign, with only a slight increase in MFE. Further, we validate the accuracy of the computational TIE metric using large-scale parallel translation experimental data. This study highlights the importance of a joint design strategy for the 5' UTR and CDS in optimizing mRNA performance, paving the way for more efficient mRNA vaccines and therapeutics.

\end{abstract}


\section{Introduction}

Messenger RNA (mRNA) vaccines and therapeutics have garnered significant attention for their potential to combat various diseases, including infectious diseases and cancer. Ensuring adequate protein translation is crucial for the efficacy of mRNA vaccines and therapies. Achieving this typically requires optimizing the mRNA's translational efficiency and stability. A conventional mRNA molecule is composed of several distinct regions: a 5' cap, a 5' untranslated region (5' UTR), a coding sequence (CDS), a 3' untranslated region (3' UTR), and a polyadenylated (Poly (A)) tail (Figure 1a). The mRNA translation process can be divided into initiation, elongation, termination, and recycling phases. Sequences within these different regions regulate various stages of the translation process, influencing the mRNA's translational efficiency, stability, and protein yield. Therefore, rational sequence design is a critical step in obtaining mRNA with desired properties. The improvement of mRNA characteristics through sequence design presents significant challenges due to the expansive design space and the complex interplay between sequence and functional properties.

Existing mRNA design strategies typically focus on optimizing only a single region of mRNA, either the UTRs or the CDS, seeking to enhance specific properties of the mRNA. For example, efforts focusing on the 5' UTR, such as Optimus 5 primer \cite{sample2019human}, Cao-RF \cite{cao2021high}, and UTR-LM \cite{chu20245}, aim primarily to enhance translation initiation efficiency (TIE). Meanwhile, designs targeting the CDS, such as LinearDesign \cite{zhang2023algorithm}, concentrate on optimizing translation elongation efficiency (TEE) and stability. However, these segmented approaches have limited ability to regulate the mRNA properties, as the properties are generally influenced by the interplay between multiple regions of an mRNA.

mRNA functions as an integrated whole, and its properties are typically not determined by the sequence of a single region alone. TIE, for instance, is influenced by both the 5' UTR and CDS sequences. Features such as hairpin structures, upstream start codons, and Kozak sequences in the 5' UTR can all impact the initiation of translation (Figure 1b). Additionally, the decoding speed of codons within the 5' leader region of the CDS, as well as the structural pairing between the 5' UTR and CDS regions, can affect the efficiency of translation initiation \cite{jackson2010mechanism, tuller2015multiple} (Figure 1c). Similarly, the stability is also determined by multiple regions of mRNA. The minimum free energy (MFE), a key indicator of mRNA stability, is influenced by the entire mRNA sequence, including both the UTRs and the CDS. Additionally, UTRs contain numerous microRNA and RNA-binding protein (RBP) binding sites that significantly impact mRNA stability \cite{fabian2010regulation, li2022rna}.

\begin{figure}[htb]
  \centering
  \includegraphics[width=\textwidth]{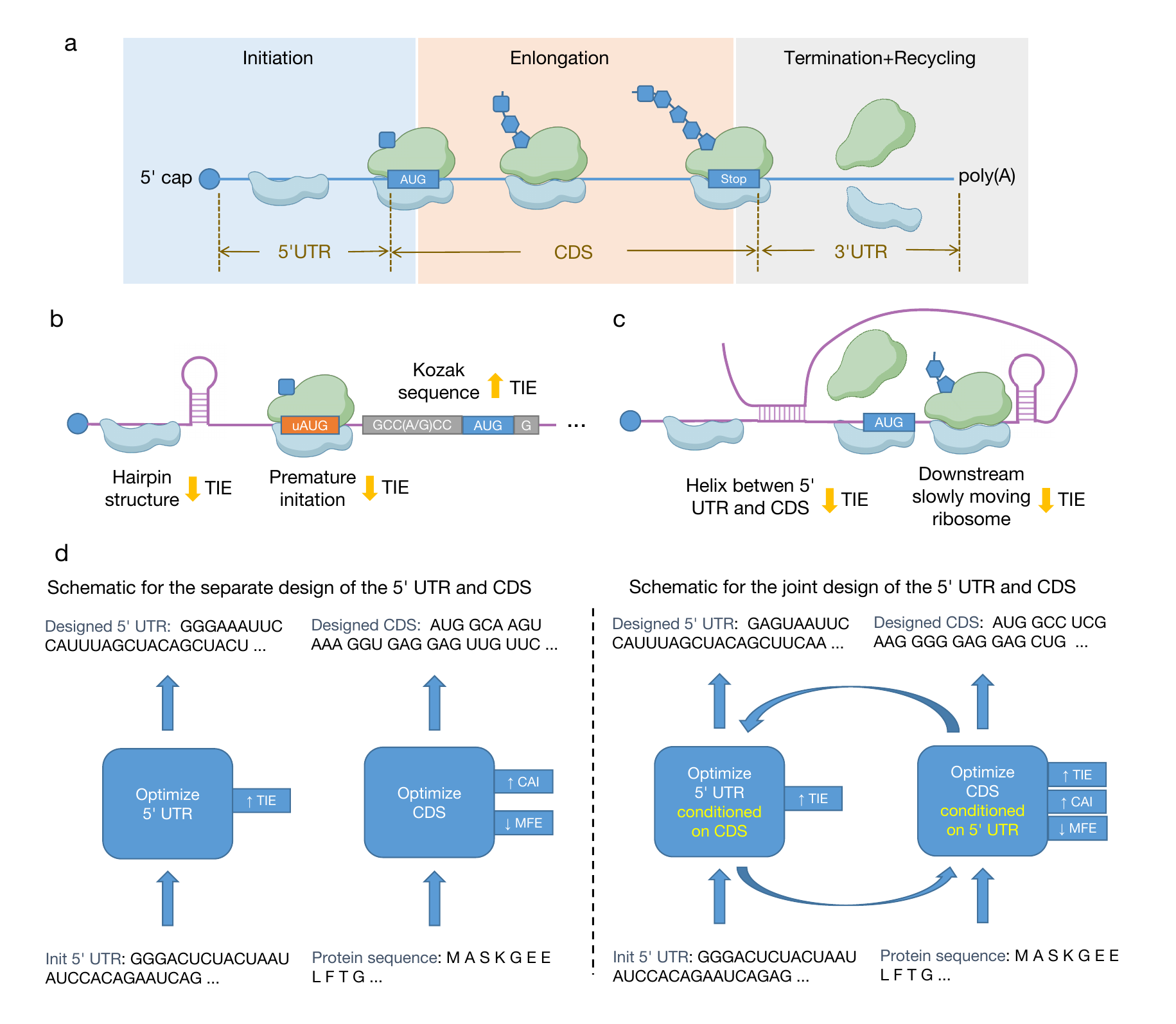}
  \caption{Schematic diagram illustrating the composition of mRNA regions, translation stages, and factors affecting its properties.
  \textbf{(a)} The composition and translation stages of mRNA. In mRNA, from the 5' end to the 3' end, the sequence includes the 5' cap, the 5' untranslated region (5' UTR), the coding sequence (CDS), the 3' untranslated region (3' UTR), and the Poly(A) tail. The CDS region is delineated by the start codon (typically AUG) at its beginning and one or more stop codons at its end. The process of mRNA translation can be divided into three stages: initiation, elongation, and termination along with recycling.
  \textbf{(b)} Factors in the 5' UTR that influence TIE. The secondary structure within the 5' UTR downregulates TIE by impeding the 40S ribosomal subunit's scanning process towards the start codon. Upstream start codons in the 5' UTR lead to premature translation initiation, thereby downregulating TIE of the CDS region. Conversely, the Kozak sequence at the 5' UTR's end promotes ribosome assembly at the start codon and enhance the TIE.
  \textbf{(c)} Factors in the CDS that influence TIE. The ribosomes located in the CDS 5' leader region spatially obstruct the assembly of subsequent ribosomes, thereby influencing the TIE of following ribosomes. Additionally, secondary structures formed between the CDS and the 5' UTR impede the 40S ribosomal subunit's scanning within the 5' UTR.
  \textbf{(d)} Schematic Illustration of 5' UTR and CDS Design Approaches. The diagram is divided into two sections by a central dashed line. The left side depicts the conventional approach of designing the 5' UTR and CDS separately, where each region is optimized in isolation. The right side illustrates the integrated design strategy employed in this study, which involves joint optimization of the 5' UTR and CDS sequences to achieve coordinated improvements in translation efficiency and stability.
}
  \label{fig:fig_1}
\end{figure}

Given the intricate interplay among diverse mRNA regions and their respective attributes, a coordinated design strategy encompassing both the UTR and CDS is imperative for striking an equilibrium between critical parameters such as TIE, TEE, and mRNA stability, ultimately enhancing the overall protein output. In this study, we introduce LinearDesign2, an advancement over the LinearDesign algorithm, which facilitates the concurrent design of the 5’ UTR and CDS, as illustrated in Figure 1d. This approach ensures a comprehensive optimization of TIE, TEE, and stability. To quantify TIE, we employ a principled, data-driven model; TEE is assessed using the Codon Adaptation Index (CAI); and mRNA stability is gauged by the MFE of the entire mRNA structure. By establishing parameterized targets, our method permits adaptable tuning of the target sequence's propensities towards these various properties, catering to the diverse needs of different biological contexts. The workflow iteratively alternates between the evolutionary optimization of the 5’ UTR conditioned on the CDS, and the optimization of the CDS conditioned on the 5’ UTR. Each region is refined with respect to a composite objective that integrates TIE, MFE, and CAI. Through iterative cycles of these optimization steps, the methodology fosters improved harmony between the 5’ UTR and CDS, leading to augmented translation efficiency and a balanced optimization across the predefined metrics.

\section{Results}
\label{sec:Results}

\subsection{TIE Contributes to the Explanation of Protein Yield in mRNAs}
LinearDesign2 expands upon the original LinearDesign algorithm by supporting the joint design of 5' UTR and CDS regions. While LinearDesign initially focused on optimizing the MFE and CAI of the CDS, it did not account for TIE. By modeling TIE base on a machine learning model, a correlation was observed: lower MFE often corresponds with reduced TIE, indicating a trade-off between these two objectives, as shown in Figure 2. Thus, when optimizing MFE, it is important to simultaneously consider TIE to prevent significant reductions. In response, LinearDesign2 integrates TIE regulation and evaluates the entire mRNA sequence’s structure and thermodynamics. Its optimization objectives include TIE, MFE, and CAI, with TIE quantified through a combined principle- and data-driven approach (see section 3.2 for details). 

To delve into the directive significance of the optimization objectives outlined in LinearDesign2, we undertake a comparative analysis of the actual protein yields obtained from samples originating from diverse regions within the metric space. Our investigation relies on the mRNA sequences and expression data sourced from the comprehensive study conducted by Leppek et al. \cite{leppek2022combinatorial}. This study meticulously quantified the expression levels of the Nluc reporter protein across a wide array of CDS sequences by utilizing the Nluc/Fluc reporter activity ratios as a metric.

Figures 2a and 2b vividly depict the expression levels of each sample at the 6-hour and 24-hour marks, respectively. In these illustrations, a darker shade of the data points signifies higher expression levels. Upon scrutinizing the distribution, it becomes evident that the samples exhibiting the peak protein expression predominantly occupy regions defined by low MFE (<-250 kcal/mol), high CAI (>0.72), and moderate TIE (0.25-0.43). This discernible trend is further corroborated by the 2D distribution plots showcased in Figures 2c and 2d.

Samples with predicted TIE values that are excessively low demonstrate a marked disadvantage in their expression levels at both time points, underscoring the necessity of adequate translation efficiency for achieving substantial protein yield. Moreover, the distribution patterns of expression levels at the 6-hour and 24-hour intervals display distinct differences. Notably, samples with the highest expression levels at the 6-hour mark tend to exhibit higher TIE, whereas those with the highest expression levels at the 24-hour mark tend to possess lower MFE. This observation aligns seamlessly with the general principle that short-term expression levels are more heavily influenced by translation efficiency, whereas long-term expression levels are more profoundly affected by mRNA stability.

However, in this distribution, the samples with the highest TIE do not exhibit high protein expression levels. This may be because these samples also have relatively high MFE values, leading to reduced stability and consequently affecting their sustained expression capability. The drawback between TIE and MFE as optimization targets is understandable since a decrease in MFE increases the compactness of mRNA structure, thereby creating obstacles for the binding of ribosome and other translation factors to mRNA \cite{gaspar2013mrna}. When filtering out samples with excessively high MFE (>-350 kcal/mol) and low CAI (<0.75), a more pronounced positive correlation between TIE and protein expression levels can be observed (Figure 2e). Specifically, in the 24-hour expression data, the Spearman correlation between TIE and Nluc/Fluc activity ratio reaches 0.73 (p < 0.05). Therefore, by ensuring relatively optimal values for MFE and CAI, optimizing TIE is likely to achieve better protein yields.

We also evaluated an additional dataset measuring protein expression levels in yeast \cite{dvir2013deciphering}. In this dataset, samples shared the same CDS and 3' UTR regions, differing only in their 5' UTR sequences. As a result, the CAI values were identical across all samples, and the MFE values showed only minor variations. Thus, we focused on assessing the explanatory power of the TIE metric for protein expression levels. As shown in Figure 2f, the Spearman correlation between the predicted TIE and the measured protein expression was 0.58 (p < 0.001), indicating that TIE can effectively account for the changes in protein expression driven by variations in the 5' UTR sequences. Overall, TIE, MFE, and CAI can be used in combination to explain protein expression changes resulting from 5' UTR or CDS modifications, providing a compelling set of optimization targets for the joint design of 5' UTR and CDS regions.


\begin{figure}[htb]
  \centering
  \includegraphics[width=\textwidth]{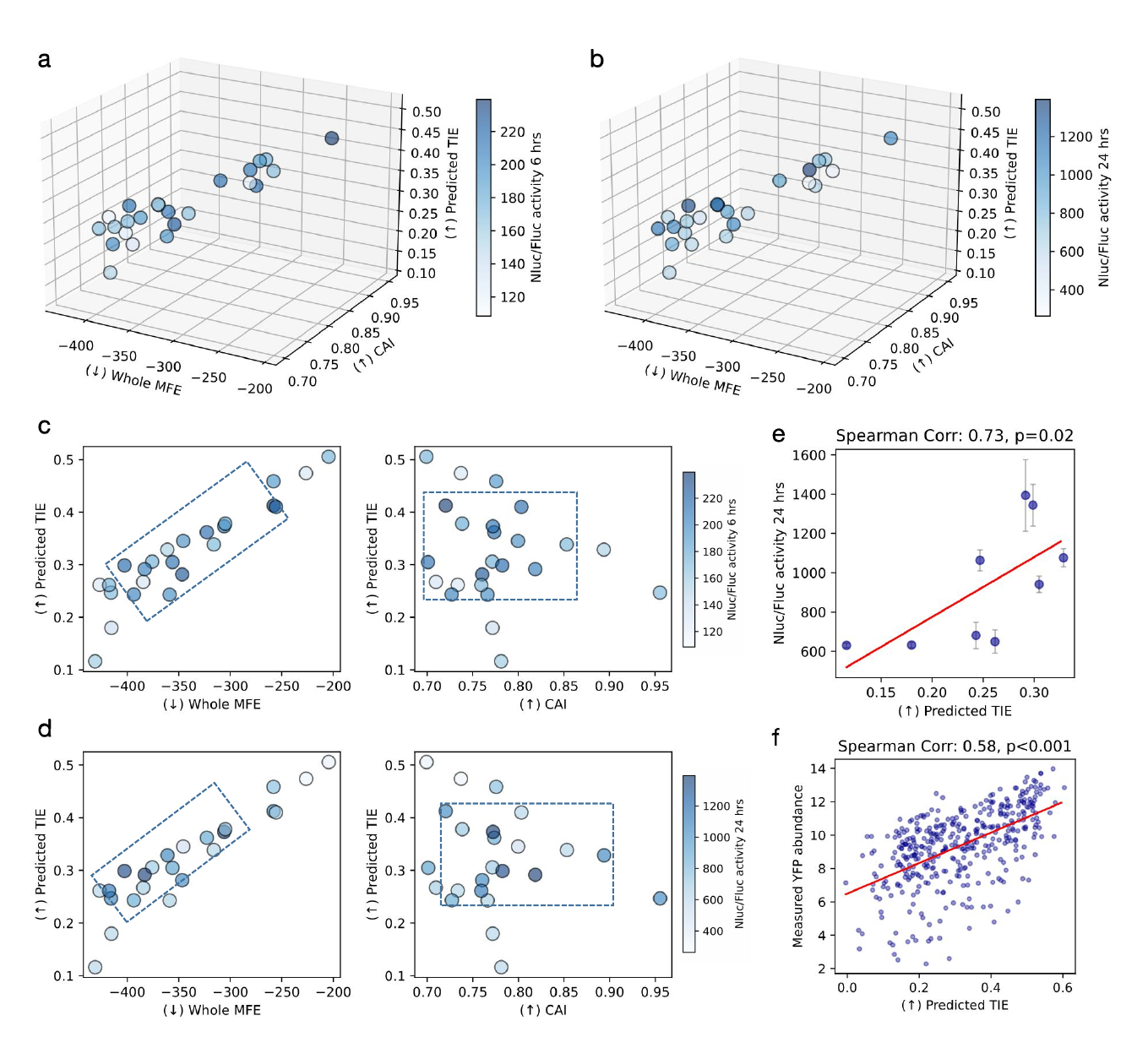}
  \caption{Distribution of mRNA sequences in metric Space and their corresponding protein expression levels.
    \textbf{(a,b)} Distribution of Nluc fluorescent protein mRNA sequences in the 3D metric space. The circle color indicates the mean of Nluc/Fluc reporter activity ratios over 6 hours \textbf{(a)} and 24 hours \textbf{(b)}.
    \textbf{(c,d)} Distribution of Nluc fluorescent protein mRNA sequences in the 2D metric space of TIE and another metric, i.e., MFE or CAI. The circle color indicates the mean of Nluc/Fluc reporter activity ratios over 6 hours \textbf{(c)} and 24 hours \textbf{(d)}. The dashed boxes indicate the area where the high-expression samples are enriched.
    \textbf{(e)} Scatters showing the correlation between TIE and Nluc/Fluc activity over 24 hours, where the samples are selected based on the criteria of MFE < -350 kcal/mol and CAI > 0.75. 
    \textbf{(f)} Scatters showing the correlation between TIE and measured YFP abundance expressed in Yeast over 24 hours.
    }
  \label{fig:fig_2}
\end{figure}

\subsection{LinearDesign2 Improves mRNA Designs in Optimizing TIE}

We assess the design outcomes of both LinearDesign and LinearDesign2 on the SARS-CoV-2 spike protein and the varicella-zoster virus (VZV) antigen. In the case of the SARS-CoV-2 spike protein (Figure 3a), the MFE values of mRNAs designed by LinearDesign are notably lower than those of the wild-type (WT) and commercial vaccine sequences (BNT-162b2 and mRNA-1273), suggesting a significant enhancement in structural compactness and thermodynamic stability. However, the TIE values of the sequences designed by LinearDesign are considerably lower, indicating a potential decrease in translation efficiency. Conversely, the sequences designed by LinearDesign2 demonstrate significantly higher TIE metrics compared to those designed by LinearDesign, while maintaining comparable CAI and MFE values.
A similar trend is observed in the design cases for the VZV antigen (Figure 3b). When compared to the wild-type (gE-WT) and a sequence designed by a codon optimization tool from Thermo Fisher \cite{raab2010geneoptimizer} (gE-Ther), the sequences designed by LinearDesign exhibit notable advantages in terms of MFE. However, they fall short in the TIE metric. LinearDesign2 successfully addresses this issue by retaining LinearDesign's advantages in MFE while significantly improving TIE performance. This suggests that LinearDesign2 can generate sequences with superior overall performance in both translation efficiency and stability. Based on the previous analysis of the relationship between the metric space and protein expression levels, this improvement is anticipated to lead to increased protein yields.

There are also noteworthy differences in the secondary structures of sequences designed by LinearDesign2 and LinearDesign. Figure 3c presents a visual comparison of the secondary structures of two eGFP mRNA sequences generated by these two algorithms. Despite both sequences exhibiting comparable metrics in terms of MFE and CAI, the sequence designed by LinearDesign2 demonstrates a significantly better TIE metric. The primary distinctions between the structures of the designed sequences are evident in the AUG-context region, encompassing both the 5' UTR and the 5' leader region of the CDS. In this region, the sequence designed by LinearDesign2 has fewer hairpin structures and base pairings, resulting in a more structurally relaxed configuration. This looser structure is generally considered advantageous for ribosome binding and scanning, thereby enhancing translation initiation efficiency. Upon extracting and folding the sequences of the AUG-context regions separately, it becomes apparent that the sequence designed by LinearDesign2 has fewer secondary structures and higher folding free energy, as shown in Figure 3d. These findings suggest that LinearDesign2 is capable of optimizing the TIE of mRNA by adjusting the secondary structure of the AUG-context region.

\begin{figure}[htbp]
  \centering
  \includegraphics[width=\textwidth]{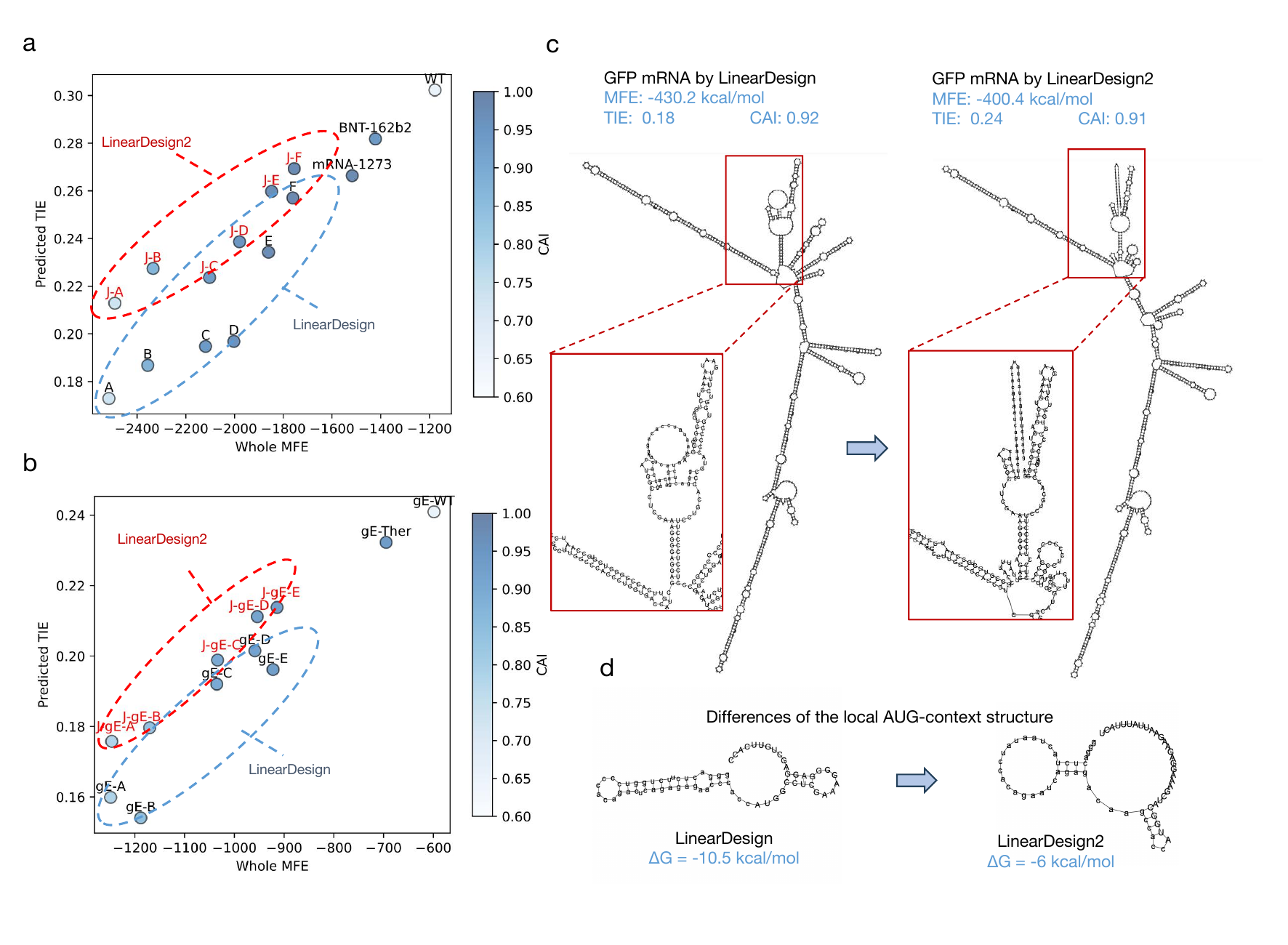}
  \caption{Comparison of metrics among wild-type, LinearDesign, LinearDesign2, and third-party designed sequences. 
    \textbf{(a)} Distribution of design outcomes for SARS-CoV-2 spike protein mRNA sequences. The horizontal axis represents overall MFE, the vertical axis represents TIE, and the color indicates CAI. Sequences designed by LinearDesign show higher structural compactness and lower TIE compared to wild-type and commercial vaccine sequences. In contrast, sequences designed by LinearDesign2 exhibit higher TIE while maintaining similar CAI and MFE metrics.
    \textbf{(b)} Distribution of design outcomes for VZV antigen mRNA vaccine sequences. Similar to the SARS-CoV-2 spike protein case, sequences designed by LinearDesign2 show improved TIE with minimal changes in MFE and CAI compared to sequences designed by LinearDesign.
    \textbf{(c)} Secondary structures of eGFP mRNA sequences designed by LinearDesign (left) and LinearDesign2 (right). The main structural differences are observed in the start codon region, where the LinearDesign2 sequence shows fewer hairpin structures and base pairings, resulting in a more relaxed configuration favorable for ribosome binding and scanning.
    \textbf{(d)} Differences in the secondary structures of the local AUG-context region for sequences designed by LinearDesign and LinearDesign2. The sequence designed by LinearDesign2 has fewer secondary structures and higher folding free energy, supporting the improved translation initiation efficiency.
    }
  \label{fig:fig_3}
\end{figure}

\subsection{Correlation between Predicted TIE and Experimental Measured Data}
In this section, we evaluate the accuracy of our TIE prediction model by comparing its predicted TIE values with experimentally measured data. In contrast to previous studies, our assessment exclusively focuses on samples whose 5' UTRs do not contain upstream AUG motifs. The presence of upstream AUGs can alter the translation initiation site and reduce the translation efficiency of the main open reading frame (ORF), which is particularly relevant in the context of mRNA vaccine development where such motifs are typically removed to optimize translation efficiency. To ensure a comprehensive evaluation, we curated and processed seven datasets from the literature, encompassing a diverse range of experimental methods (including polysome profiling, ribosome profiling, and fluorescence-activated cell sorting (FACS)), protein types (such as eGFP, mCherry, YFP, and various proteins from the human genome), and species (including human and yeast). This diversity enables us to robustly validate the generalizability of our model. The details of data processing and the train-test splits are provided in section 3.3.

Figure 4 presents the performance of our TIE prediction model, compared to representative models in the field. Specifically, Figure 4a shows the results of models trained on the TE dataset of endogenous mRNA from the PC3 cell line (Human\_genome) and evaluated on various test sets. This test setting is challenging, as recent studies indicate that existing TIE prediction models exhibit low generalizability across different datasets \cite{schlusser2024current}. Spearman correlation coefficients between the predicted TIE values and experimental measurements are used as the performance metric. We developed three versions of our model based on different input features: Ours-UTR, which uses only the 5' UTR sequence; Ours-UTR-CDS, which includes both the 5' UTR sequence and the 5' leader portion of the CDS; and Ours-UTR-CDS-Auxfeats, which further incorporates knowledge-driven features including the GC content and structural stability around AUG, as well as overall mRNA structural stability derived from both 5' UTR and CDS sequences. The construction details of these models are described in section 3.2.

Across all test sets, the model that incorporated both the 5' UTR and the 5' leader region of the CDS (Ours-UTR-CDS) consistently outperformed the model that used only the 5' UTR as input (Ours-UTR). This result underscores the added value of incorporating CDS sequence information for TIE prediction. Furthermore, the inclusion of knowledge-driven features in the Ours-UTR-CDS-Auxfeats model led to significant additional improvements in accuracy. When compared to other representative models, such as Optimus5primer \cite{sample2019human} and UTR-LM \cite{chu20245}, our model demonstrated clear advantages. These benchmark models primarily rely on 5' UTR sequences alone. In contrast, our model leverages both the 5' UTR and CDS sequences, as well as domain knowledge-driven features, which collectively contributed to its superior predictive performance.

Given the relatively small sample size of the Human\_genome dataset, we anticipated potential limitations in the model's performance. To address this, we developed a TIE prediction model trained on multi-source data. Since the datasets were obtained using different experimental methods, the annotations varied in physical interpretation and scale, making multi-source model training non-trivial. We devised tailored training strategies to handle this complexity, as detailed in section 3.2. After incorporating multi-source data, the model's performance further improved across all test sets, as shown in Figure 4b. For comparison, since existing models do not support multi-source training, we trained Optimus5primer and UTR-LM separately on two datasets (Human\_genome and eGFP\_75nt). We observed that, in most cases, our model trained on multi-source data significantly outperformed the models trained on a single dataset. One exception occurred when Optimus5primer trained on the eGFP\_75nt dataset slightly outperformed our model on the eGFP\_75nt test set, likely due to the similarity in data distribution between the training and test sets.

Overall, the improvement in TIE prediction accuracy achieved by our model can be attributed to the comprehensive utilization of both 5' UTR and CDS sequences, along with domain knowledge-driven features, and the integration of multi-source data. These results underscore the importance of jointly optimizing 5' UTR and CDS regions in mRNA sequence design to enhance translation initiation efficiency.

\begin{figure}[htbp]
  \centering
  \includegraphics[width=0.8\textwidth]{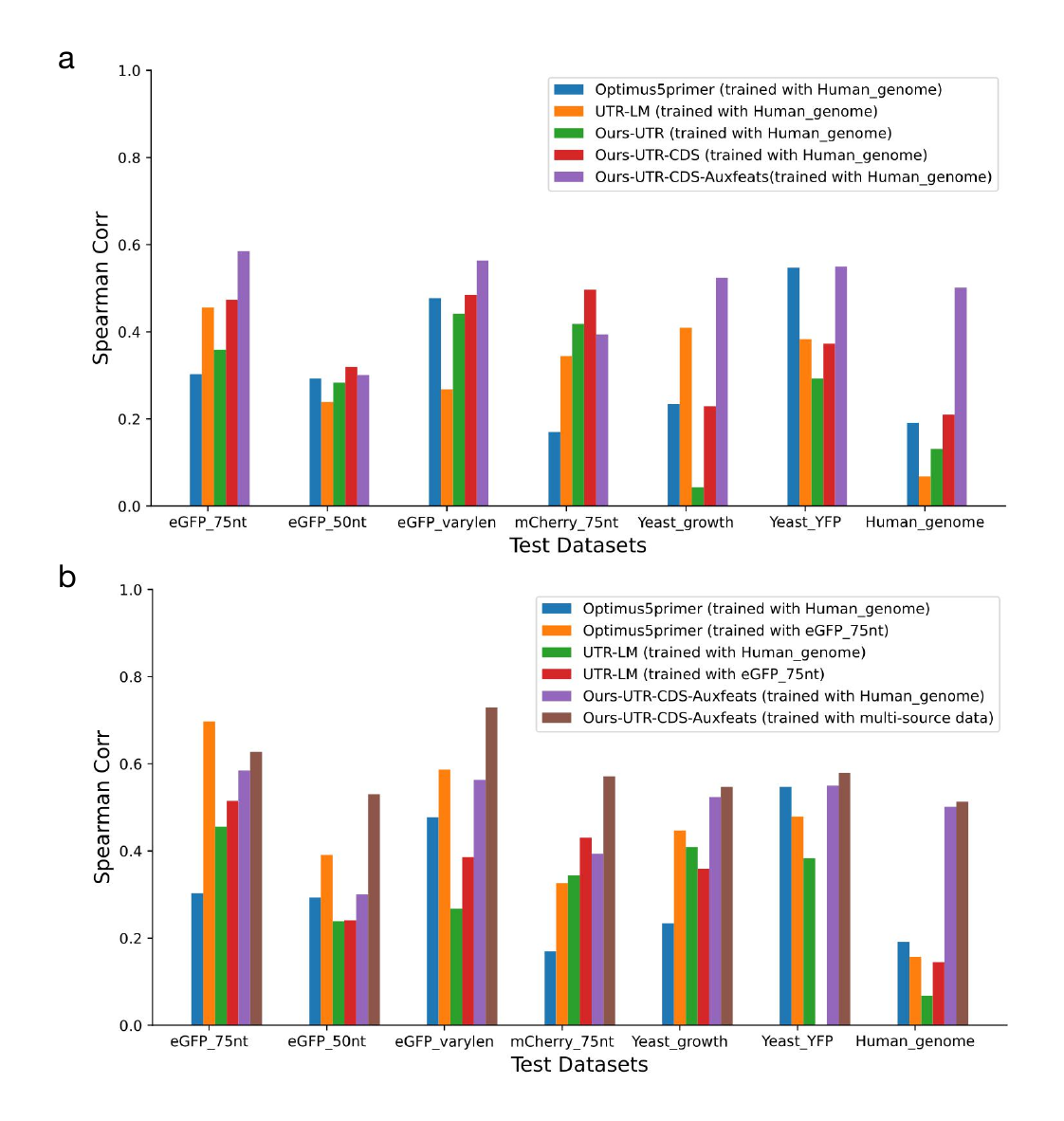}
  \caption{TIE prediction performance of our TIE predictor and other representative models in the field. 
    \textbf{(a)} Performance of models trained on the TE dataset of endogenous mRNA from the PC3 cell line (Human\_genome). 
    \textbf{(b)} Performance of our TIE predictor trained on multi-source data, and comparison with models trained on single dataset.
    }
  \label{fig:fig_5}
\end{figure}

\section{Method}
\subsection{Optimization Target and Process of LinearDesign2}

The optimization target of the LinearDesign2 algorithm is a combination of TIE, CAI, and MFE. TIE is a measure of how effectively ribosomes initiate translation on an mRNA molecule. By optimizing TIE, we aim to enhance the overall efficiency of protein synthesis, ensuring a robust and rapid response when the mRNA is delivered into cells. CAI is a metric that reflects the usage of synonymous codons in a given mRNA sequence. CAI is related to the translational elongation efficiency of a gene based on the codon usage bias. Optimizing CAI ensures that the mRNA uses codons that are preferred by the host's translational machinery, which can enhance the rate and accuracy of protein synthesis. MFE is a measure of the stability of the mRNA secondary structure. A lower MFE indicates a more stable structure, which can protect the mRNA from degradation and improve its longevity in the cellular environment. However, overly stable structures might impede ribosome binding and translation initiation. Therefore, MFE must be balanced with TIE and CAI to achieve optimal mRNA performance.

The optimization target combines these three metrics using two balancing factors, $\lambda_{TIE}$ and $\lambda_{CAI}$, to coordinate their relative weights:

\begin{equation}
S = \lambda_{TIE} L \cdot \log(TIE) + \lambda_{CAI} L \cdot \log(CAI) - MFE
\label{eq:optimization_target}
\end{equation}

Here, $L$ represents the number of codons in the CDS region. The inclusion of $L$ with TIE and CAI terms ensures that these terms are on a comparable numerical scale to MFE. The logarithmic transformation of TIE and CAI (denoted as $\log(TIE)$ and $\log(CAI)$) converts multiplicative interactions among internal factors into additive ones, simplifying the calculation. 

$\lambda_{TIE}$ controls the weight of TIE in the optimization target. By adjusting $\lambda_{TIE}$, the algorithm can prioritize translation initiation efficiency relative to the other metrics. $\lambda_{CAI}$ controls the weight of CAI in the optimization target. Adjusting $\lambda_{CAI}$ allows the algorithm to emphasize codon usage optimization as needed. When $\lambda_{TIE}$ is set to 0, the optimization target reduces to the original LinearDesign, focusing solely on MFE and CAI. This flexibility enables the algorithm to be tailored to different optimization needs, balancing translation efficiency, codon usage, and structural stability to achieve the desired mRNA properties.

\begin{figure}[htbp]
  \centering
  \includegraphics[width=1.0\textwidth]{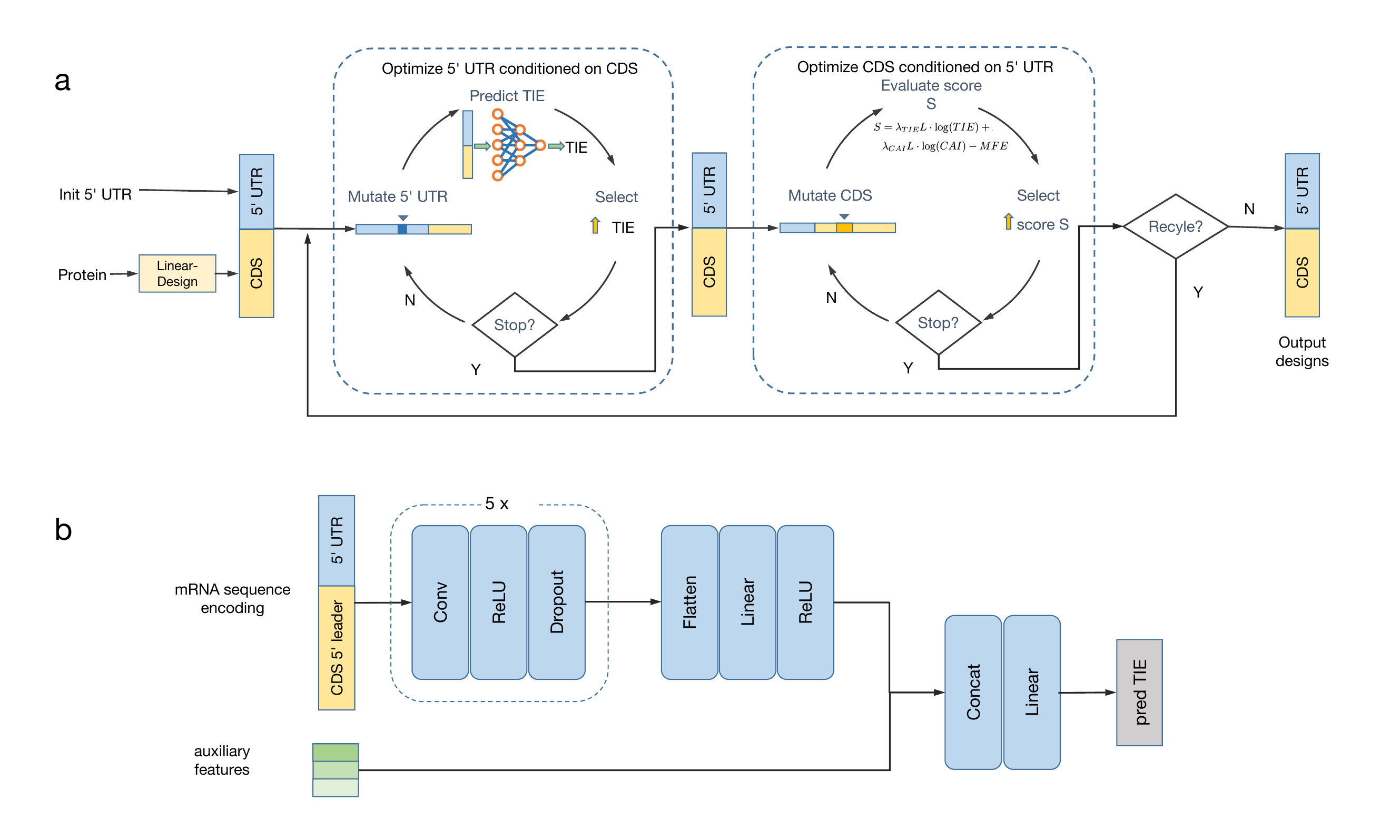}
  \caption{The process of LinearDesign2 and the network structure of TIE prediction model.
    \textbf{(a)} The Pipeline of LinearDesign2 for joint design of 5' UTR and CDS.
    \textbf{(b)} The network architecture for predicting the TIE of an mRNA sequence.
    }
  \label{fig:fig_6}
\end{figure}

The process of joint design for the 5' untranslated region (5'UTR) and coding sequence (CDS) begins with input consisting of the target protein's amino acid sequence and an initial 5'UTR sequence. It is preferable to use an initial 5'UTR sequence known to support normal protein expression. The workflow is composed of four main steps, as shown in Figure 5a, combining sequence optimization and evolutionary algorithms to improve the overall optimization target as defined above.

Step 1, initial CDS design using LinearDesign. In the first step, the initial CDS sequence is designed based on the provided amino acid sequence using the LinearDesign algorithm. This process optimizes both the CAI and MFE, with $\lambda_{CAI}$ being utilized to balance the weights of these two metrics during the optimization procedure.

Step 2, 5'UTR optimization under fixed CDS conditions. In the second step, the 5'UTR sequence is  optimized while keeping the CDS sequence fixed. An evolutionary algorithm is employed to perform random mutations in the 5'UTR sequence, generating an initial population of N 5'UTR variants. Each variant is combined with the CDS sequence to predict the TIE using our TIE prediction model. The bottom 50\% of the population, based on the TIE score, are filtered out, and the remaining top-performing variants undergo further rounds of mutation, evaluation, and selection. This iterative process continues until a predefined number of iterations is reached, at which point the 5'UTR with the highest TIE score is selected for the next step.

Step 3, CDS optimization under fixed 5'UTR conditions. Once the 5'UTR sequence is optimized, the focus shifts to optimizing the CDS sequence. Again, an evolutionary algorithm is applied, but this time synonymous codon substitutions are performed in the CDS region. For each CDS variant, CAI and MFE are recalculated, and the variant is combined with the 5'UTR for TIE prediction. Variants that exceed a certain threshold of CAI change are discarded to ensure that CAI remains stable throughout the optimization. Each variant's overall score $S$ is calculated using Equation 1, incorporating CAI, MFE, and TIE scores. The overall scores are then used to drive the selection process. After several rounds of optimization, the CDS variant with the highest overall score is chosen.

Step 4, recycling Steps 2 and 3 for further optimization. In the final step, the process recycles Steps 2 and 3 iteratively. By alternating between optimizing the 5'UTR sequence (Step 2) and the CDS sequence (Step 3), this recycling approach allows for further coordination between the 5'UTR and CDS designs, leading to improved optimization of the target metrics. The number of recycling iterations needs to balance effectiveness with computational efficiency. Typically, 1–2 cycles of recycling are sufficient to achieve significant enhancements in the design outcomes.


\subsection{Quantification of Translation Initiation Efficiency (TIE)}

We developed a TIE prediction model based on a deep neural network (DNN) architecture. The input to the model includes the 5' UTR sequence and the first 30 nucleotides (nt) from the 5' end of the CDS, allowing for a comprehensive approach to modeling the effects of both the 5' UTR and CDS sequences on TIE. Additionally, several domain-specific handcrafted features were introduced into the input to further enhance the model’s generalization ability. These features are listed in Table 1.

Our TIE prediction model follows a convolutional neural network (CNN) architecture, as illustrated in Figure 5b. For the sequence input, we employ a 5-layer 1D CNN to extract features. Each convolutional layer contains 32 filters with a kernel size of 9 and a dilation rate of 2 to expand the receptive field. Every convolutional layer is followed by a ReLU activation layer and a Dropout layer, with the dropout rate set to 0.2. The output from the CNN is passed through a Flatten layer to convert it into a feature vector. To accommodate variable-length inputs, we apply flattening only to the tensor positions corresponding to the last 60 nt of the 5' UTR. The flattened feature vector is then processed by a linear layer combined with ReLU to reduce its dimensionality to 32, after which it is concatenated with the handcrafted auxiliary features. Finally, the combined vector is passed through a linear layer to predict TIE.

Training a model using multi-source data presents several challenges due to the differences in experimental methods, label types, and data scales across datasets. These differences make it difficult to directly compare samples from different datasets, preventing a straightforward combination of data for training. Furthermore, the available labels, such as mean ribosome load (MRL), translation efficiency (TE), and protein expression levels, are indirect reflections of TIE rather than direct measurements, introducing noise into the data. To address this, we propose a training strategy based on intra-source ranking. During training, all samples within a batch must originate from the same data source. Batches from different sources are collected separately, avoiding the issue of incomparability between datasets. Within each batch, we construct pairs of samples for ranking. In each pair, the difference in the labeled values between the samples must exceed a threshold $t$, reducing the impact of label noise on the training process. A margin ranking loss function is used to calculate the ranking loss for each pair of samples, as defined by the following formula:

\begin{equation}
\mathcal{L}_{rank} = \max(0, m - y(s_i - s_j))
\label{eq:loss_function}
\end{equation}

\begin{equation}
y =
\begin{cases}
    1, & \text{if } y_i - y_j > t \\
    -1, & \text{if } y_i - y_j < -t \\
    0, & \text{otherwise}
\end{cases}
\end{equation}

where $y_i$ and $y_j$ are the ground-truth labels of the two samples, $s_i$ and $s_j$ are the predicted scores, and $m$ is the margin value.

\begin{table}[ht]
\centering
\caption{Domain Knowledge-Driven Features Incorporated into the TIE Prediction Model}
\label{tab:TIE_features}
\begin{tabular}{Sc p{10cm}}
\toprule
\textbf{Feature Name} & \textbf{Description} \\ 
\midrule
MFE\_per\_nucleotide & The MFE of the mRNA, normalized by the sequence length \\ 
AUG\_context\_ddG & $MFE_{unfold}(M) - MFE(M)$,  where $MFE_{unfold}(M)$ is the MFE of mRNA $M$ with the AUG context (including 5’ UTR and CDS 5’ leader region) constrained to be unpaired, and $MFE(M)$ is the MFE of mRNA $M$ with no structure constraints. \\ 
AUG\_context\_GC\_content & GC contents in the AUG context region. \\
\bottomrule
\end{tabular}
\end{table}

\subsection{Dataset Collection and Processing}
We collected a total of seven datasets for training and evaluating our model. The details of these datasets are presented in Table 2. These datasets were derived from various studies and employed different experimental techniques, including polysome profiling, ribosome profiling, and fluorescence-activated cell sorting (FACS). Four datasets—eGFP\_75nt, eGFP\_50nt, eGFP\_varylen, and mCherry\_75nt—were sourced from the work of Seelig and colleagues \cite{sample2019human, castillo2024optimizing}. These datasets contain mean ribosome load (MRL) labels measured in human HEK293T cells using polysome profiling. Since the CDS and 3' UTR regions in these experiments were fixed sequences, MRL captures the impact of 5' UTR variations on translation initiation efficiency (TIE).

The Yeast\_growth dataset, also from a work of Seelig’s team \cite{cuperus2017deep}, uses yeast growth rate measurements to infer the protein yield of the HIS3 reporter gene containing different 5' UTR variants. As the 5' UTR primarily affects translation initiation, the protein yield serves as an indirect indicator of changes in TIE. The Yeast\_YFP dataset, collected by Dvir et al. \cite{dvir2013deciphering}, employs FACS to measure changes in YFP protein yield caused by alterations in the 5' UTR, similarly providing an indirect measure of TIE changes. 

These datasets above all utilize a single type of CDS for the reporter, limiting the ability to assess the impact of CDS sequences on TIE. To address this, we incorporated the Human\_genome dataset. The Human\_genome dataset, derived from the work of Hsieh and colleagues \cite{hsieh2012translational}, uses ribosome profiling to measure translation efficiency (TE) for each endogenous mRNA transcript in the human PC3 cell line. Since translation initiation is often considered the rate-limiting step of endogenous mRNA translation efficiency \cite{shah2013rate}, TE measured via ribosome profiling can serve as an indicator of TIE.

We performed extensive cleaning of the raw data from these sources, filtering out low-confidence samples with too less reads. Since our focus is on TIE prediction for samples without upstream AUG (uAUG), we removed samples containing uAUG from all datasets except for Human\_genome. Retaining uAUG-containing samples in Human\_genome is intended to preserve the diversity of CDS sequences in the dataset. We then split each dataset into training and testing sets. For datasets with a single type of CDS, we imposed an identity constraint between the training and test sets, ensuring that sequence identity was less than 0.7. For the Human\_genome dataset, we ensured that gene types in the training and test sets did not overlap and included all uAUG-containing samples in the training set. The number of samples in the training and test sets for each dataset after processing is shown in Table 2.

\begin{table}[htbp]
\centering
\caption{Summary of Datasets Used for TIE Prediction}
\resizebox{\textwidth}{!}{
\begin{tabular}{Slp{1.5 cm}p{1.5 cm}p{3 cm}p{2.5 cm}p{1.5 cm}p{1.3 cm}}
\toprule
     \textbf{Dataset} &        \textbf{Label type} & \textbf{Species, cell type} &                \textbf{5’ UTR format} &                  \textbf{CDS types} &  \textbf{Training samples} &  \textbf{Test samples} \\
\midrule
   eGFP\_75nt\cite{sample2019human} &               MRL &     Human, HEK293T &     fixed 25nt + random 50nt &                    1, eGFP &           24190 &        5810 \\
   eGFP\_50nt\cite{castillo2024optimizing} &               MRL &     Human, HEK293T &                  random 50nt &                    1, eGFP &           11056 &        2764 \\
eGFP\_varylen\cite{sample2019human} &               MRL &     Human, HEK293T & fixed 25nt + random 25\textasciitilde 100nt &                    1, eGFP &           14451 &        3679 \\
mCherry\_75nt\cite{sample2019human} &               MRL &     Human, HEK293T &     fixed 25nt + random 50nt &                 1, mCherry &           26178 &        3822 \\
Yeast\_growth\cite{cuperus2017deep} &       growth rate &              Yeast &                  random 50nt &                    1, HIS3 &            8312 &        2078 \\
   Yeast\_YFP\cite{dvir2013deciphering} & protein abundance &              Yeast &      Fixed 7nt + Random 10nt &                     1, YFP &            1591 &         398 \\
Human\_genome\cite{hsieh2012translational} &                TE &         Human, PC3 &       human endogenous 5’UTR & 4562, human endogenous CDS &            4128 &         434 \\
\bottomrule
\end{tabular}
}
\label{tab:datasets}
\end{table}

\section{Conclusions}

In this study, we introduced the LinearDesign2 algorithm, a novel approach aimed at the joint optimization of the 5' untranslated region (5' UTR) and coding sequence (CDS) in mRNA design. Through comprehensive computational experiments conducted on various antigens, including SARS-CoV-2 and Varicella-Zoster Virus (VZV), we demonstrated that LinearDesign2 achieves significant improvements across multiple key metrics: Codon Adaptation Index (CAI), Minimum Free Energy (MFE), and Translation Initiation Efficiency (TIE). Notably, the advancements in TIE are particularly pronounced compared to the original LinearDesign, holding promise for substantially enhancing mRNA translation efficiency and, consequently, protein yield.

To bolster the computational optimization of TIE, we developed a new TIE prediction model. This model innovatively incorporates joint modeling of 5' UTR and CDS sequences, enriched with domain knowledge-driven features. This integration allows the model to capture the intricate interplay between 5' UTR and CDS, thereby providing more accurate TIE predictions. Furthermore, by refining the training methodology on diverse datasets, we enhanced both the precision and generalization capability of our TIE prediction model.

The LinearDesign2 algorithm, powered by this sophisticated TIE prediction model, offers a more holistic and balanced solution for mRNA design. By addressing the intricate relationships among the sequence, structure and properties of mRNAs and focusing on critical factors influencing translation efficiency, LinearDesign2 is poised to unlock greater protein expression potential. This, in turn, could accelerate the development of mRNA-based vaccines and other mRNA-related therapies, bringing us closer to realizing the full potential of mRNA technology in medical applications.

\bibliographystyle{unsrt}  
\bibliography{references}

\end{document}